\documentstyle[prd,aps,epsfig,eqsecnum,preprint,tighten]{revtex}

\bibstyle{unsrt}
\begin{document}
\draft


\title{Solution of Schwinger-Dyson Equations for ${\cal PT}$-Symmetric Quantum
Field Theory}

\author{Carl M. Bender,$^1$,\thanks{electronic mail: cmb@howdy.wustl.edu}
Kimball A. Milton,$^2$\thanks{electronic mail: milton@mail.nhn.ou.edu}
 and Van M. Savage$^1$\thanks{electronic mail: vmsavage@hbar.wustl.edu}
 }
\address{${}^1$Department of Physics, Washington University, St. Louis, Missouri
63130, USA}
\address{${}^2$Department of Physics and Astronomy, University of Oklahoma,
Norman, Oklahoma 73019, USA}

\date{\today}
\preprint{OKHEP--99--05}
\maketitle
 
\begin{abstract}
In recent papers it has been observed that non-Hermitian Hamiltonians, such as
those describing 
$ig\phi^3$ and $-g\phi^4$ field theories, still possess real positive spectra so
long as the weaker condition of ${\cal PT}$ symmetry holds. This allows for the
possibility of new kinds of quantum field theories that have strange and quite
unexpected properties. In this paper a technique based on truncating the
Schwinger-Dyson equations is presented for renormalizing and solving such field
theories. Using this technique it is argued that a $-g\phi^4$ scalar quantum
field theory in four-dimensional space-time
is renormalizable, is asymptotically free, has a
nonzero value of $\langle0|\phi|0\rangle$, and has a positive definite spectrum.
Such a theory might be useful in describing the Higgs boson.
\end{abstract}

]
 
\section{Introduction}
\label{s1}

It has recently been observed \cite{r1,r2} that quantum mechanical theories
whose Hamiltonians are ${\cal PT}$-symmetric have positive definite spectra even
if the Hamiltonian is not Hermitian. A class of such theories that has been
studied extensively is defined by the Hamiltonian
\begin{equation}
H=p^2-(ix)^N\quad(N\geq2).
\label{eq1.1}
\end{equation}
It is believed that the reality and positivity of the spectra are a direct
consequence of ${\cal PT}$ symmetry.

The positivity of the spectra for all $N$ is an extremely surprising result; it
is not at all obvious, for example, that the Hamiltonian $H=p^2-x^4$
corresponding to $N=4$ has a positive real spectrum. To understand this result
it is necessary to define properly the boundary conditions in the corresponding
Schr\"odinger equation.

For the Hamiltonian in Eq.~(\ref{eq1.1}) the Schr\"odinger differential equation
corresponding to the eigenvalue problem $H\psi=E\psi$ is 
\begin{equation}
-\psi''(x)-(ix)^N\psi(x)=E\psi(x).
\label{eq1.2}
\end{equation}
The boundary conditions for this equation are discussed in detail in
Ref.~\cite{r1}. There, it was shown how to continue analytically in the
parameter $N$ away from the harmonic oscillator value $N=2$. This analytic
continuation defines the boundary conditions in the complex-$x$ plane. The
regions in the cut complex-$x$ plane in which $\psi(x)$ vanishes exponentially
as $|x|\to\infty$ are {\it wedges}. In Ref.~\cite{r1} the wedges for $N>2$ were
chosen to be analytic continuations of the wedges for the harmonic oscillator,
which are centered about the negative and positive real axes and have angular
opening ${\pi\over2}$. For arbitrary $N>2$ the anti-Stokes' lines at the centers
of the left and right wedges lie below the real axis at the angles
\begin{eqnarray}
\theta_{\rm left}&=&-\pi+\left({N-2\over N+2}\right){\pi\over2},\nonumber\\
\theta_{\rm right}&=&-\left({N-2\over N+2}\right){\pi\over 2}.
\label{eq1.3}
\end{eqnarray}
The opening angle of these wedges is ${2\pi\over N+2}$. In Ref.~\cite{r1} the
time-independent Schr\"odinger equation was integrated numerically inside the
wedges to determine the eigenvalues to high precision.

The quantum mechanical Hamiltonian in Eq.~(\ref{eq1.1}) has additional
remarkable properties. For example, for all $N>2$ the expectation value
$\langle0|x|0\rangle$ of the position operator $x$ in the ground state
is nonzero. This surprising result is true even when $N=4$ \cite{r1}.

These results for quantum mechanics raise some interesting questions regarding
quantum field theory. In particular, does the self-interacting scalar quantum
field theory defined by the Lagrangian
\begin{equation}
{\cal L}={1\over2}(\partial\phi)^2+{1\over2}m^2\phi^2-{g\over N}(i\phi)^N
\label{eq1.4}
\end{equation}
have a positive definite spectrum and a nonvanishing value of $\langle0|\phi|0
\rangle$ for all $N>2$? We believe that the answer to this question is yes.
Because of these properties, we believe that when $N=4$ the resulting quantum
field theory in four-dimensional space-time could serve as a good description of
the Higgs particle. As we argue in this paper, the $-g\phi^4$ theory is
particularly advantageous because, like the conventional $g\phi^4$ it has a
dimensionless coupling constant, but unlike the conventional theory, it is
asymptotically free and is thus not a trivial theory.

The question of how to determine the properties of ${\cal PT}$-symmetric
non-Hermitian quantum field theories has already been examined. As we will see
in this paper, conventional Feynman diagrammatic perturbation theory is not
adequate for studying these theories. Thus, in previous work \cite{r3,r4} the
perturbative approach that was used was to take $N=2+\delta$, where $\delta$ is
treated as a small parameter. While some interesting results regarding parity
violation \cite{r3} and supersymmetry \cite{r4} were obtained, unfortunately
this perturbative scheme has a severe drawback: It is not known how to carry out
the renormalization procedure required to understand higher-dimensional
theories.

Why is it that Feynman diagrams cannot be used to perform calculations in
theories such as $-g\phi^4$? As we have already stated, in this theory $\langle0
|\phi|0\rangle\neq0$. There is no way to obtain this result using the standard
Feynman rules; one cannot obtain a one-point Green's function using four-point
vertex amplitudes. Indeed, as we will show in Sec.~\ref{s3} in the context
of zero-dimensional theories, the standard Feynman rules are incorrect for this
theory.

In this paper we perform a systematic truncation of the
Schwinger-Dyson equations as a calculational procedure. This idea has already
been applied in a simple context to obtain Green's functions and energy levels
in conventional quantum mechanical problems \cite{r5}. Truncating the
Schwinger-Dyson equations is an inherently variational approach; including more
and more of the higher Green's functions is equivalent to enlarging the space of
variational parameters. In a recent study of the ${\cal PT}$-symmetric,
non-Hermitian quantum mechanical Hamiltonian in Eq.~(\ref{eq1.1}) variational
methods were found to be extremely accurate \cite{r6}.

This paper is organized as follows: In Sec.~\ref{s2} we review the general
approach used in this paper. We show how to derive the Schwinger-Dyson equations
using simple functional methods and we explain our truncation procedure. In
Sec.~\ref{s3} we examine the numerical accuracy of our truncation method in the
context of zero-dimensional field theory. We study the massless zero-dimensional
version of Eq.~(\ref{eq1.4}) in very high order in the truncation process for
the cases $N=3$ and $N=4$. In the case of a zero-dimensional massive theory
we show that the Feynman rules are inapplicable.
In Sec.~\ref{s4} we examine the theory in
Eq.~(\ref{eq1.4}) in one-dimensional space-time (quantum mechanics). Finally, in
Sec.~\ref{s5} we apply our methods to field theory in arbitrary dimension. As an
application of our procedure we calculate the Callan-Symanzik function
$\beta(g)$ for a
four-dimensional $-g\phi^4$ theory for small $g$. We show that to leading order
$\beta(g)$ is negative and thus the theory is asymptotically free. This result
is in distinct contrast with the result for a conventional $g\phi^4$ theory.

\section{Elementary Derivation of the Schwinger-Dyson Equations}
\label{s2}

The Schwinger-Dyson equations are an infinite set of coupled equations relating
the Green's functions of a quantum field theory. In this section we derive the
Schwinger-Dyson equations using elementary formal functional methods.

We begin with Eq.~(\ref{eq1.4}) and append a term that represents the coupling
of the field $\phi(x)$ to an external $c$-number source $J(x)$:
\begin{equation}
{\cal L}={1\over2}(\partial\phi)^2+{1\over2}m^2\phi^2-{g\over N}(i\phi)^N-J\phi.
\label{eq2.1}
\end{equation}
This Lagrangian represents a self-interacting scalar quantum field theory in
$D$-dimensional Euclidean space-time. If we vary the action with respect to
$\phi(x)$ we obtain the field equation:
\begin{equation}
-\partial^2\phi(x)+m^2\phi(x)-ig[i\phi(x)]^{N-1}=J(x).
\label{eq2.2}
\end{equation}

Next, leaving the source turned on, we take the expectation value of the field
equation (\ref{eq2.2}) in the vacuum state of the theory $|0\rangle$ and divide
by the vacuum-vacuum functional $Z[J]=\langle0|0\rangle$:
\begin{eqnarray}
-\partial^2G_1^{(J)}(x)+m^2G_1^{(J)}(x)
-gi^N{\langle0|\phi^{N-1}(x)|0\rangle
\over\langle0|0\rangle}=J(x),
\label{eq2.3}
\end{eqnarray}
where $G_1^{(J)}(x)$ is the one-point Green's function in the presence of the
external source:
\begin{equation}
G_1^{(J)}(x)\equiv{\langle0|\phi(x)|0\rangle\over\langle0|0\rangle}.
\label{eq2.4}
\end{equation}
Note that the function $J(x)$ appears alone on the right side of
Eq.~(\ref{eq2.3}) because it is a $c$-number and therefore can be factored
out of matrix elements.

The objective is now to use Eq.~(\ref{eq2.3}) to calculate the Green's functions
of the theory. The (connected) Green's functions in the presence of the source
$J$ are defined as functional derivatives of the logarithm of $Z[J]$ with
respect to the source $J(x)$:
\begin{eqnarray}
G_n^{(J)}(x_1,x_2,\dots,x_n)
\equiv{\delta^n\over\delta J(x_1)\delta J(x_2)\cdots\delta J(x_n)}\ln(Z[J]).
\label{eq2.5}
\end{eqnarray}
To obtain the standard connected Green's functions of the theory (the connected
part of the vacuum expectation value of the time-ordered product of the fields;
that is, the sum of the connected $n$-point Feynman diagrams) we then turn off
the source:
\begin{equation}
G_n(x_1,x_2,\dots,x_n)=G_n^{(J)}(x_1,x_2,\dots,x_n)\Bigm|_{J\equiv0}.
\label{eq2.6}
\end{equation}
Note that turning off the source restores translation invariance. As a result,
the one-point Green's function $G_1$ is a constant independent of $x$, the
two-point Green's function depends only on the difference $x-y$, $G_2(x,y)=G_2(
x-y)$, $G_3(x,y,z)=G_3(x-y,x-z)$ depends on two differences, and so on.

Before one can proceed, one must express the third term in Eq.~(\ref{eq2.3}) in
terms of the connected Green's functions of the theory. To do this we recall
that functionally differentiating with respect to $J(x)$ is equivalent to
inserting $\phi(x)$ in matrix elements \cite{r5}.

Let us consider the simple case $N=3$. In this case it is necessary to
calculate the quantity $\langle0|\phi^2(x)|0\rangle$. To do so we begin with
Eq.~(\ref{eq2.4}) multiplied by $Z[J]$:
\begin{equation}
G_1^{(J)}(x)Z[J]=\langle0|\phi(x)|0\rangle.
\label{eq2.7}
\end{equation}
Taking the functional derivative of this equation with respect to $J(x)$ gives
\begin{equation}
[G_1^{(J)}(x)]^2 Z[J]+G_2^{(J)}(x,x) Z[J]=\langle0|\phi^2(x)|0\rangle.
\label{eq2.8}
\end{equation}
Hence, we can eliminate $\langle0|\phi^2(x)|0\rangle$ from Eq.~(\ref{eq2.3}) to
obtain
\begin{eqnarray}
-\partial^2G_1^{(J)}(x)+m^2G_1^{(J)}(x)
+gi\left([G_1^{(J)}(x)]^2+G_2^{(J)}(x,x)\right)=J(x).
\label{eq2.9}
\end{eqnarray}
We now obtain the first of the Schwinger-Dyson equations by setting $J\equiv0$
(turning off the source):
\begin{equation}
m^2G_1+gi\left[G_1^2+G_2(0)\right]=0.
\label{eq2.10}
\end{equation}
Remember that by translation invariance $G_1$ is a constant, so that its
derivative vanishes and that $G_2(0)=G_2(x-x)=G_2(x,x)$.

To obtain the second of the Schwinger-Dyson equations for $N=3$ we take a
functional derivative of Eq.~(\ref{eq2.9}) with respect to $J(y)$,
\begin{eqnarray}
-\partial^2G_2^{(J)}(x,y)+m^2G_2^{(J)}(x,y)+gi\left[2G_1^{(J)}(x)G_2^{(J)}
(x,y)
+G_3^{(J)}(x,x,y)\right]=\delta(x-y),
\label{eq2.11}
\end{eqnarray}
and then set $J\equiv0$ in this equation:
\begin{eqnarray}
-\partial^2G_2(x-y)+m^2G_2(x-y)
+gi\left[2G_1G_2(x-y)+G_3(0,x-y)\right]=\delta(x-y).
\label{eq2.12}
\end{eqnarray}

If we continue the process of functionally differentiating with respect to $J$
and setting $J\equiv0$, we obtain the infinite tower of coupled differential
equations known as
the Schwinger-Dyson equations. For example, the third in the sequence is
\begin{eqnarray}
&&-\partial^2G_3(x-y,x-z)+m^2G_3(x-y,x-z)\nonumber\\
&&\qquad +gi[2G_1G_3(x-y,x-z)+2G_2(x-z)G_2(x-y)
+G_4(0,x-y,x-z)]=0.
\label{eq2.13}
\end{eqnarray}

Note that these Schwinger-Dyson equations are {\it incomplete} in the sense that
there are too many unknowns. The first equation contains $G_1$ and $G_2$, the
second contains $G_1$, $G_2$, and $G_3$, and so on. Thus, each new equation
contains a new unknown Green's function and the system never closes. However, we
can force the system to close by truncating the sequence of coupled equations
and setting $G_{n+1}=0$ in the last equation. We will use this method throughout
the remainder of the paper.

As a second example we derive the first four Schwinger-Dyson Equations from
Eq.~(\ref{eq2.3}) for the case $N=4$. Using the same approach as we did for the
case $N=3$, we begin by reexpressing $\langle0|\phi^3(x)|0\rangle$. We do this
by taking the functional derivative of Eq.~(\ref{eq2.8}) with respect to $J(x)$
to obtain 
\begin{eqnarray}
[G_1^{(J)}(x)]^3 Z[J]+3G_1^{(J)}(x)G_2^{(J)}(x,x)Z[J]
+G_3^{(J)}(x,x,x) Z[J]=\langle0|\phi^3(x)|0\rangle.
\label{eq2.14}
\end{eqnarray}
Substituting this result into the $N=4$ version of Eq.~(\ref{eq2.3}), we have
\begin{eqnarray}
-\partial^2G_1^{(J)}(x)+m^2G_1^{(J)}(x)-g\left([G_1^{(J)}(x)]^3
+3G_1^{(J)}(x)G_2^{(J)}(x,x)+G_3^{(J)}(x,x,x)\right)=J(x).
\label{eq2.15}
\end{eqnarray}
We obtain the first Schwinger-Dyson equation for $N=4$ by setting $J\equiv0$:
\begin{equation}
m^2G_1-g\left[G_1^3+3G_1G_2(0)+G_3(0,0)\right]=0.
\label{eq2.16}
\end{equation}

Using the same procedure as for $N=3$ we now take a functional derivative of
Eq.~(\ref{eq2.15}) with respect to $J(y)$ to obtain
\begin{eqnarray}
&&-\partial^2G_2^{(J)}(x,y)+m^2G_2^{(J)}(x,y)-g\left(3 [G_1^{(J)}(x)]^2G_2^{(J)}
(x,y)\right.\nonumber\\
&&\quad+3G_1^{(J)}(x)G_3^{(J)}(x,x,y)+3G_2^{(J)}(x,x)G_2^{(J)}(x,y)
 \left.+G_4^{(J)}(x,x,x,y)\right)=\delta(x-y).
\label{eq2.17}
\end{eqnarray}
Now, setting $J\equiv0$ gives the second of the Schwinger-Dyson equations for
$N=4$:
\begin{eqnarray}
&&-\partial^2G_2(x-y)+m^2G_2(x-y) -g[ 3 G_1^2 G_2(x-y) \nonumber\\
&&\qquad +3 G_2(0) G_2(x-y)+3G_1G_3(0,x-y)
+G_4(0,0,x-y)]=\delta(x-y).
\label{eq2.18}
\end{eqnarray}

Repeating this process once more by functionally differentiating
Eq.~(\ref{eq2.17}) with respect to $J(z)$ and setting $J\equiv0$ gives 
\begin{eqnarray}
&&-\partial^2G_3(x-y,x-z)+m^2G_3(x-y,x-z)\nonumber\\
&&\qquad-g[6G_1G_2(x-y)G_2(x-z) +3 G_1^2 G_3(x-y,x-z)\nonumber\\
&&\qquad+3G_2(x-z)G_3(0,x-y)
+3G_2(x-y)G_3(0,x-z)+3G_2(0)G_3(x-y,x-z)\nonumber\\
&&\qquad +3G_1G_4(0,x-y,x-z)+G_5(0,0,x-y,x-z)]=0,
\label{eq2.19}
\end{eqnarray}
the third of the Schwinger-Dyson equations for $N=4$.
The fourth Schwinger-Dyson equation is
\begin{eqnarray}
&&-\partial^2G_4(x-y,x-z,x-w)+m^2G_4(x-y,x-z,x-w)\nonumber\\
&&\qquad-g[6G_2(x-y)G_2(x-z) G_2(x-w)
+6G_1G_2(x-y)G_3(x-z,x-w)\nonumber\\
&&\qquad+6G_1G_2(x-z)G_3(x-y,x-w)
+6G_1G_2(x-w)G_3(x-y,x-z)\nonumber\\
&&\qquad+3G_3(0,x-y)G_3(x-z,x-w)
+3G_3(0,x-z)G_3(x-y,x-w)\nonumber\\
&&\qquad+3G_3(0,x-w)G_3(x-y,x-z)
+3G_1^2G_4(x-y,x-z,x-w)\nonumber\\
&&\qquad+3G_2(x-y)G_4(0,x-z,x-w)
+3G_2(x-z)G_4(0,x-y,x-w)\nonumber\\
&&\qquad+3G_2(x-w)G_4(0,x-y,x-z)
+3G_2(0)G_4(x-y,x-z,x-w)\nonumber\\
&&\qquad+3G_1G_5(0,x-y,x-z,x-w)
+G_6(0,0,x-y,x-z,x-w)]=0.
\label{eq2.20}
\end{eqnarray}

Again, we observe that the set of equations is incomplete; for this case the
number of unknown Green's functions is two more than the number of equations
(instead of one more as in the case $N=3$). That is, the first equation for
$N=4$ contains the three unknowns $G_1$, $G_2$, and $G_3$, the second contains
$G_1$, $G_2$, $G_3$, and $G_4$, and so on. To solve this system of equations we
truncate after the $n$th equation, but now to close the system of equations we
must set $G_{n+1}=0$ and $G_{n+2}=0$ in the last and next to last equations. 

How do we generalize this derivation to arbitrary $N$? Clearly, when $N$ is an
{\it integer} and $N>2$ we can use the two cases $N=3$ and $N=4$ discussed above
as paradigms. All that is needed is to functionally differentiate
Eq.~(\ref{eq2.7})
$N-2$ times. This gives an expression for $\langle0|\phi^{N-1}(x)|0\rangle$ in
Eq.~(\ref{eq2.3}). For example, when $N=5$ we have
\begin{eqnarray}
&&Z[J]\left([G_1^{(J)}(x)]^4+6[G_1^{(J)}(x)]^2 G_2^{(J)}(x,x)
+3[G_2^{(J)}(x,x)]^2 \right.\nonumber\\
&&\qquad +4 G_1^{(J)}(x) G_3^{(J)}(x,x,x)
\left.+G_4^{(J)}(x,x,x,x)\right) =\langle0|\phi^4(x)|0\rangle,
\label{eq2.21}
\end{eqnarray}
when $N=6$, we have
\begin{eqnarray}
&& Z[J]\left([G_1^{(J)}(x)]^5+10 [G_1^{(J)}(x)]^3 G_2^{(J)}(x,x)
+15 G_1^{(J)}(x) [G_2^{(J)}(x,x)]^2\right.\nonumber\\
&&\quad+10 [G_1^{(J)}(x)]^2 G_3^{(J)}(x,x,x)
+10 G_2^{(J)}(x,x) G_3^{(J)}(x,x,x)
\nonumber\\
&&\quad \left.+5 G_1^{(J)}(x)G_4^{(J)}(x,x,x,x)
+G_5^{(J)}(x,x,x,x,x)\right)=\langle0|\phi^5(x)|0\rangle,
\label{eq2.22}
\end{eqnarray}
and so on. Once this calculation is completed, repeated functional
differentiation with respect to $J$ followed by setting $J\equiv0$ gives the
complete set of coupled Green's function equations.

Note that while these equations are rather complicated, they are easily
expressible in terms of multinomial coefficients. Following the notation of
Abramowitz and Stegun \cite{r7}, the multinomial coefficients are defined as
follows: For each integer $n$, there is a set of multinomial coefficients; each
coefficient expresses the number of possible ways to partition $n=a_1+2a_2+
\ldots+na_n$ different objects into $a_k$ subsets containing $k$ objects
($k=1,~2,~\ldots,~n$). For the first few integers $n$ the sets of multinomial
coefficients (called $M_3$ in Ref.~\cite{r7}) are $\{1\}$ for $n=1$, $\{1,1\}$
for $n=2$, $\{1,3,1\}$ for $n=3$, $\{1,4,3,6,1\}$ for $n=4$, and $\{1,5,10,10,
15,10,1\}$ for $n=5$. Observe that these are precisely the coefficients that
appear in (\ref{eq2.7}) for $\langle0|\phi(x)|0\rangle$, (\ref{eq2.8}) for
$\langle0|\phi^2(x)|0\rangle$, (\ref{eq2.14}) for $\langle0|\phi^3(x)|0\rangle$,
(\ref{eq2.21}) for $\langle0|\phi^4(x)|0\rangle$, and (\ref{eq2.22}) for
$\langle0|\phi^5(x)|0\rangle$. Also, in these equations the powers of the
Green's functions are the numbers $a_k$.

To be precise, for each $n$ we must take all possible combinations of the
numbers $a_k$ satisfying $n=a_1+2a_2+\ldots+na_n$. In the notation of
Ref.~\cite{r7}, $\pi=1^{a_1},2^{a_2},\dots,n^{a_n}$. This allows us to read off
the subscripts of the Green's functions and the powers to which they are raised;
the numbers $a_k$ are the powers and the numbers they exponentiate are the
subscripts of the Green's functions in that term. From this we calculate the
quantity
\begin{eqnarray}
M_3=(n;a_1,a_2,\dots,a_n)
={n!\over(1!)^{a_1}a_1!(2!)^{a_2}a_2!\dots(n!)^{a_n}a_n!},
\label{eq2.23}
\end{eqnarray}
which is the coefficient for that particular combination of Green's functions.

For example, given $\langle0|\phi^3(x)|0\rangle$ there are three possible
combinations of the numbers $a_k$ that satisfy $n=a_1+2a_2+\ldots+na_n$:
$a_1=3$; $a_1=1$ and $a_2=1$; and $a_3=1$.
This gives $\pi=1^3$; $\pi=1,2$; and $\pi=3$. Thus, there are terms of the form
$G_1^3$, $G_1G_2$, and $G_3$. Calculating $M_3$ we find that the coefficients of
these terms are 1, 3, and 1. Hence, we are able to reconstruct
Eq.~(\ref{eq2.14}). 

When $N$ is noninteger, the situation is much more difficult. To derive an
expression for $\langle0|\phi^{N-1}(x)|0\rangle$ in this case, we assume first
that $N$ is integer and use the general formula for $\langle0|\phi^{N-1}(x)|0
\rangle$ in terms of multinomial coefficients. Then, in principle, we can
continue analytically off the integers using analytical expressions in terms
of Gamma functions for these multinomial coefficients.

Of course, this procedure is complicated, but we illustrate it below by
considering the simplest truncation possible in which we keep only the first two
Schwinger-Dyson equations and set $G_n=0$ for all $n>2$. In this case, for
integer values of $N$ we have the beautiful result that the only multinomial
coefficients that appear are precisely the coefficients of the Hermite
polynomials. Thus, for {\sl noninteger} values of $N$ we have parabolic cylinder
functions with the exponential divided off. It is in this fashion that we are
able to continue off the integers $n$.

To be explicit, we use the standard notation $D_\nu(x)$ to represent the
parabolic cylinder function \cite{r8} and define the function $w_\nu(x)$ by
\begin{equation}
D_\nu(x)=e^{-x^2/4}x^\nu w_\nu(x).
\label{eq2.24}
\end{equation}
Then, for integer $N$ we factor $[G_1^{(J)}]^N$ out of the equation for $\langle
0|\phi^N(x)|0\rangle$ and introduce the variables $\gamma^{(J)}$ and $\gamma$ by
\begin{equation}
\gamma^{(J)}={iG_1^{(J)}(x)\over\sqrt{G_2^{(J)}(x,x)}}\quad{\rm and}\quad
\gamma={iG_1\over\sqrt{G_2(0)}}.
\label{eq2.25}
\end{equation}
We then obtain for arbitrary noninteger $N$
\begin{eqnarray}
{\langle0|\phi^N(x)|0\rangle\over Z[J]}
&=& [G_1^{(J)}(x)]^N w_N(\gamma^{(J)})\nonumber\\
&=&[G_1^{(J)}(x)]^N\sum_{k=0}^\infty{(-1)^kN!\over(N-2k)! 2^k k!
[\gamma^{(J)}]^{2k}}.
\label{eq2.26}
\end{eqnarray} 
(This series terminates if $N$ is integer but is an infinite series if $N$ is
noninteger.) As we will see, when $J\equiv0$, the constant $G_1$ is a negative
imaginary number and the constant $G_2(0)$ is real and positive. Thus, the
argument $\gamma$ of $w_N$ is real and positive.

Using Eq.~(\ref{eq2.26}) we write the field equation (\ref{eq2.3}) explicitly in
terms of the Green's functions of the theory:
\begin{eqnarray}
-\partial^2G_1^{(J)}(x)+m^2G_1^{(J)}(x)
-gi^N [G_1^{(J)}(x)]^{N-1} w_{N-1}(\gamma^{(J)})=J(x).
\label{eq2.27}
\end{eqnarray}  
(Remember that in the derivation of this equation we have discarded all
Green's functions higher than $G_2$.) As before, we obtain the first of the
Green's functions for arbitrary $N$ by setting the source $J\equiv0$:
\begin{equation}
m^2G_1-gi^N G_1^{N-1}w_{N-1}(\gamma)=0.
\label{eq2.28}
\end{equation}		

We obtain the second of the Schwinger-Dyson equations for arbitrary $N$ by
taking the functional derivative of Eq.~(\ref{eq2.27}) with respect to $J(y)$:
\begin{eqnarray}
&&-\partial^2G_2^{(J)}(x,y)+m^2G_2^{(J)}(x,y)\nonumber\\
&&\qquad-gi^N\Bigg((N-1)[G_1^{(J)}(x)]^{N-2}G_2^{(J)}(x,y)w_{N-1}(\gamma^{(J)})
\nonumber\\
&&\qquad\qquad +i[G_1^{(J)}(x)]^{N-1}w_{N-1}'(\gamma^{(J)}){G_2^{(J)}(x,y)\over
\sqrt{G_2^{(J)}(x,x)}}\Bigg)
=\delta(x-y),
\label{eq2.29}
\end{eqnarray}  
where we have set $G_3^{(J)}=0$. Now, setting $J\equiv0$ in the above equation
yields the second of the Schwinger-Dyson equations:
\begin{eqnarray}
&&-\partial^2G_2(x-y)+m^2G_2(x-y)\nonumber\\
&&\qquad-gi^N\left[(N-1) G_1^{N-2} G_2(x-y)w_{N-1}(\gamma)
+iG_1^{N-1}w_{N-1}'(\gamma){G_2(x-y)\over
\sqrt{G_2(0)}}\right]=\delta(x-y).
\label{eq2.30}
\end{eqnarray}  

This equation can be simplified by using the recurrence relations for parabolic
cylinder functions \cite{r8},
\begin{equation}
x w_{N-1}'(x)=(N-1)[w_{N-2}(x)-w_{N-1}(x)],
\label{eq2.31}
\end{equation}
to obtain
\begin{eqnarray}
-\partial^2G_2(x-y)+m^2G_2(x-y)
+(N-1)g(iG_1)^{N-2}w_{N-2}(\gamma)]G_2(x-y)=\delta(x-y).
\label{eq2.32}
\end{eqnarray}
Thus, Eq.~(\ref{eq2.28}) and Eq.~(\ref{eq2.32}) are a closed system of two
equations and two unknowns. The solution of this system for various choices of
space-time dimension will be discussed in the following sections.

\section{Zero-Dimensional Theories}
\label{s3}

In zero-dimensional space-time the integral representation for the
vacuum-vacuum functional $Z[J]$ corresponding to the Lagrangian in
Eq.~(\ref{eq1.4}) is an ordinary integral
\begin{equation}
Z[J]=\int_{-\infty}^{\infty} d\phi\,\exp\left[-{1\over2}m^2\phi^2
+{g\over N}(i\phi)^N+J\phi\right].
\label{eq3.1}
\end{equation}
To demonstrate the numerical accuracy of our truncation method we first study
massless theories of the form in Eq.~(\ref{eq3.1}) where all
quantities are exactly calculable. We find that for arbitrary $N$ the Green's
functions can be expressed exactly in terms of Gamma functions. Then, we
examine massive theories and show that weak-coupling diagrammatic methods are
inadequate for the analysis of ${\cal PT}$-symmetric theories.

\subsection{Massless Theories in Zero Dimensions}
\label{ss3a}

Using Eqs.~(\ref{eq3.1}), (\ref{eq2.5}), and (\ref{eq2.6}) the expectation value
of the field (the one-point connected Green's function) is 
\begin{eqnarray}
G_1=-i\left({4N\over g}\right)^{1/N}{\Gamma\left({1\over N}+{1\over2}\right)
\cos\left({\pi\over N}\right)\over\sqrt{\pi}},
\label{eq3.2}
\end{eqnarray}
and the expectation value of the field squared is 
\begin{equation}
{\langle0|\phi^2|0\rangle\over\langle0|0\rangle}=
\left({N\over g}\right)^{2/N}{\Gamma\left({3\over N}\right)
\left[\sin^2\left({\pi\over N}\right)-3\cos^2\left({\pi\over N}\right)\right]
\over\Gamma\left({1\over N}\right)}.
\label{eq3.3}
\end{equation} 
Using Eq.~(\ref{eq2.8}) with $J\equiv0$ we express the two-point connected
Green's function as $G_2=\langle0|\phi^2(x)|0\rangle/\langle0|0\rangle-G_1^2$,
which can be calculated exactly using the two equations given above. 

Observe that for any $N\ge2$, $G_1$ is pure imaginary and negative and $G_2$ is real
and positive, as claimed earlier. This is in fact true in any dimension. The
reality of $G_2$ and the imaginarity of $G_1$ follows from the ${\cal PT}$
symmetry of these quantities. Applying ${\cal P}$ to $G_1$ changes the sign,
while applying ${\cal P}$ to $G_2$ leaves the sign intact. Hence, under
${\cal T}$, which acts as complex conjugation, $G_1$ changes sign again and
thus is pure imaginary, while $G_2$ does not change sign and thus is real. 

The first truncation approximations to $G_1$ and $G_2$ obtained by keeping just
the first two Schwinger-Dyson equations are found by solving the
zero-dimensional massless versions of Eqs.~(\ref{eq2.28}) and (\ref{eq2.32}).
Observe that with $m=0$, Eq.~(\ref{eq2.28}) demands that
\begin{equation}
w_{N-1}(\gamma)=0.
\label{eq3.4}
\end{equation}
In zero-dimensional space-time there are no derivatives with respect to $x$ and
the delta function is unity. Thus, all Schwinger-Dyson equations are algebraic.
The second Schwinger-Dyson equation is
$$(N-1)g(iG_1)^{N-2}w_{N-2}(\gamma_0)G_2=1,$$
where $\gamma_0$ is the solution of Eq.~(\ref{eq3.4}). (Note that there is some
ambiguity with regard to which zero to choose. Our numerical studies suggest
that the most positive zero is always the correct choice, but we do not have a
proof. We find that it is this zero that gives the most accurate numerical
results.) Recalling the definition of $\gamma$, we insert $G_2=-G_1^2/
\gamma_0^2$ to obtain an expression for $G_1$:
\begin{equation}
G_1=-i\left[{\gamma_0^2\over(N-1)gw_{N-2}(\gamma_0)}\right]^{1/N}.
\label{eq3.5}
\end{equation}
Again using the definition of $\gamma$, we express $G_2$ as
\begin{equation}
G_2=\left[{\gamma_0^{2-N}\over(N-1)gw_{N-2}(\gamma_0)}\right]^{2/N}.
\label{eq3.6}
\end{equation}
Table \ref{t1} compares the exact results with the corresponding first
approximations for $N=3$, $N=4$, and $N=5$. 

Deriving an expression for the second approximation in terms of $N$ is
difficult. However, for a specific $N$, solving larger and larger systems of
Schwinger-Dyson equations in which more and more Green's functions are included
can be done symbolically on a computer. The case $N=3$ is comparatively simple.
If we keep $n$ equations, then for $k<n$, the $k$th equation is linear in
$G_{k+1}$. Thus, we can systematically solve a sequence of linear equations for
the Green's functions, substitute the results into the $n$th equation, and solve
the resulting polynomial equation for $G_1$.

For $N=3$, the first six approximations to $iG_1g^{1/3}$ are $0.79370$,
$0.69336$, $0.74690$, $0.71256$, $0.73987$, and $0.71237$. The exact answer,
taken from Eq.~(\ref{eq3.2}) with $N=3$, is
$0.72901$ (see Table \ref{t1}). Note that the approximations oscillate around
the correct answer. For the case of a Hermitian Hamiltonian we would expect
that the approach to the correct answer is monotone. This is because the
calculation that we are performing is variational in nature. Keeping more
and more Green's functions corresponds to enlarging the parameter space.
However, in this case the Hamiltonian is not Hermitian, so this oscillation
is not surprising. Indeed, it is consistent with our variational studies of
${\cal PT}$-symmetric quantum mechanical systems in Ref.~\cite{r6}.

\begin{figure}[p]
\centerline{\epsfig{file=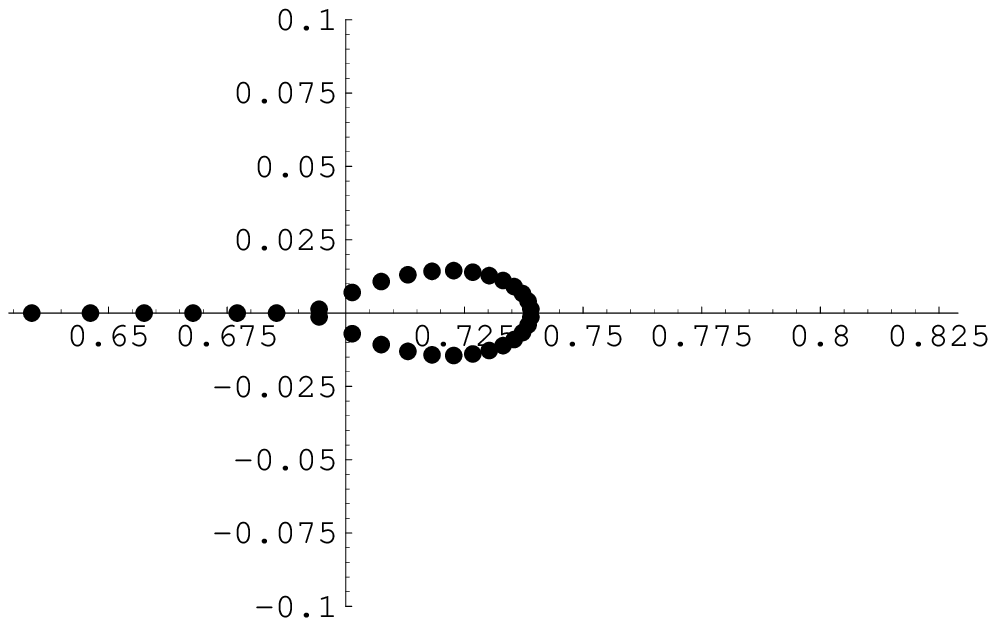,width=10cm}}
\caption{Solutions of the first 150 coupled Schwinger-Dyson equations for the
dimensionless one-point Green's function $iG_1g^{1/3}$ for the case of a
massless zero-dimensional $N=3$ theory. Note that the solutions (indicated by
dots) lie in a small portion of the complex plane very close to the exact
answer $0.72901$.}
\label{fig1}
\end{figure}

Apart from the oscillations, we would hope that successive approximations would
come closer to the correct value. Thus, the fact that the sixth approximation is
worse than the fifth and that eventually most of the approximations become
complex is a startling result. We believe that this is also due to
the non-Hermiticity of the Hamiltonian. Indeed, while these approximations
are numerically extremely accurate, we find that the convergence is rather
slow and it may even be that this truncation method does not converge. To
examine the convergence we have numerically solved {\it very large} systems of
coupled Schwinger-Dyson equations for $N=3$; we have plotted in Fig.~\ref{fig1}
all solutions for the one-point Green's function in the complex plane for
the system of 150 coupled Schwinger-Dyson equations that lie near the
exact answer. In this figure we see that the algebraic solutions for
the one-point Green's function form a very small {\it loop} around the exact
value. The largest dimension of the loop is along the real axis and is
approximately $0.04$. The furthest distance of a point on the loop from
the correct answer is approximately $0.034$. A weighted average of the points on
the loop is extremely close to the correct answer $0.72901$.

\begin{figure}[p]
\centerline{\epsfig{file=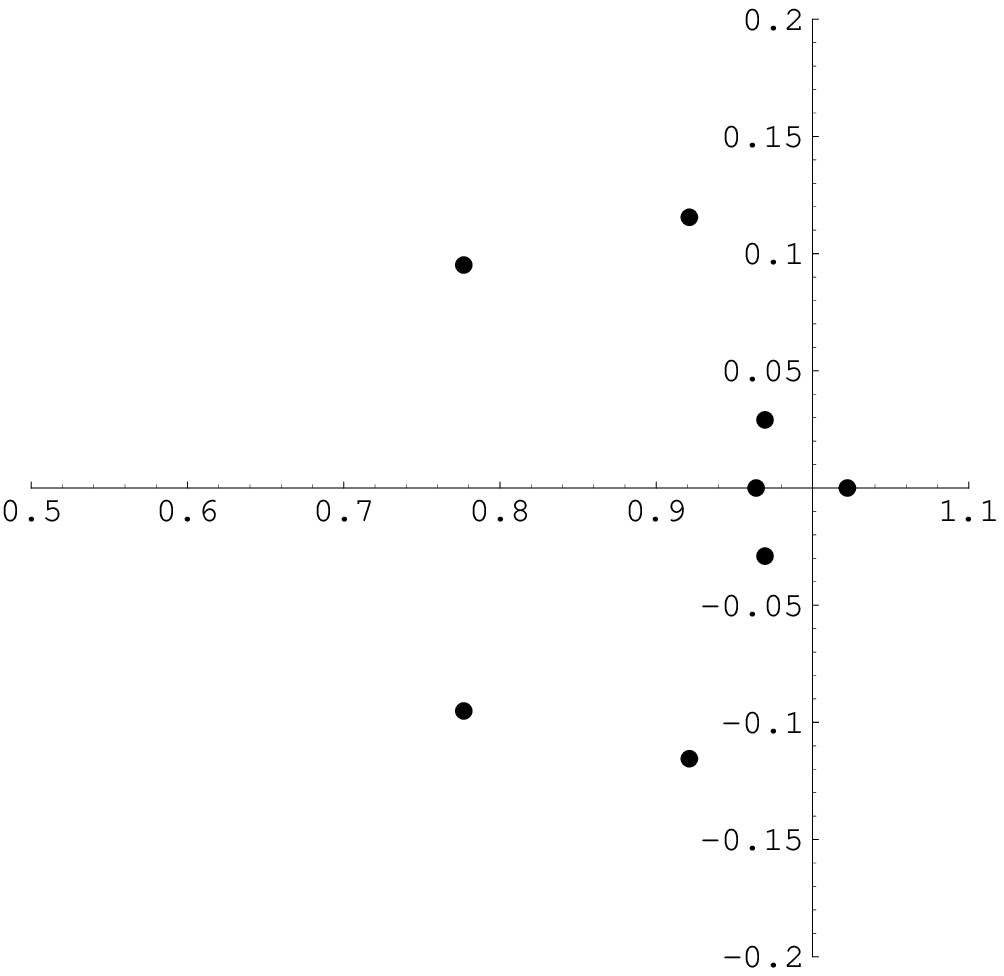,width=10cm}}
\caption{Solutions of the first ten coupled Schwinger-Dyson equations for the
dimensionless one-point Green's function $iG_1g^{1/4}$ for the case of a
massless zero-dimensional $N=4$ theory. The solutions (indicated by
dots) lie in a small portion of the complex plane very close to the exact
answer $0.97774$.}
\label{fig2}
\end{figure}

The case $N=4$ is more complicated than that of $N=3$ because now the last two 
Schwinger-Dyson equations, rather than just the last equation, are nonlinear.
As a result the largest number of equations we can solve and hence, the largest
number of Green's functions we can include, is ten. The first five
approximations to $iG_1g^{1/4}$ are $1.10668$, $0.90560$, $1.02988$, $0.96159$,
and $1.02868$. The exact value is $0.97774$ (see Table \ref{t1}). Just as
for $N=3$, the approximations oscillate around the exact answer and successive 
approximations are numerically accurate but appear to converge very slowly.
In this case, the fifth approximation is worse than the fourth. We have plotted
in Fig.~\ref{fig2} all solutions of the first ten coupled Schwinger-Dyson
equations for the one-point Green's function that lie near the exact answer.
As in the case of Fig.~\ref{fig1} these solutions form a
small loop in the complex plane around the exact answer.

The numerical work that we have done on zero-dimensional massless theories
suggests that solving systems of truncated Schwinger-Dyson equations
gives extremely accurate numerical approximations to the Green's functions. The
convergence of the method is still not understood and warrants further study.

\subsection{Massive Theories in Zero Dimensions}
\label{ss3b}

In zero-dimensional space-time, massless theories are so simple that it is
possible to solve the Schwinger-Dyson equations as algebraic systems. However,
in higher dimensions, the coupled Schwinger-Dyson equations are quite
complicated. Thus, we will be interested in obtaining perturbative solutions
for the case of small coupling constant $g$.

To gain a rudimentary understanding of the perturbative nature of the theories
in Eq.~(\ref{eq1.4}), we study in this subsection the path-integral expressions
for the $n$-point {\it disconnected} Green's functions $F_n$ of a
zero-dimensional $-g\phi^4$ field theory. These disconnected Green's functions
are expressed as ratios of integrals:
\begin{equation}
F_n={\int_{-\infty}^{\infty}d\phi\,\phi^n\exp\left(-{1\over2}m^2\phi^2
+{g\over4}\phi^4\right)\over\int_{-\infty}^{\infty} d\phi \,
\exp\left(-{1\over2}m^2\phi^2+{g\over4}\phi^4\right)}.
\label{eq3.7}
\end{equation} 

Although these integrals cannot be done exactly, we can use the method of
steepest descents to obtain an asymptotic result for small $g$ with $m$
fixed. We will see that, depending on whether the sign of $m^2$ is positive
or negative, we obtain drastically different results.

We begin by replacing $\phi$ by the dimensional quantity
$x$ according to $\phi=\sqrt{2/g}|m|x$. This gives 
\begin{equation}
F_n=\left({2|m|^2\over g}\right)^{n/2}{\int_{-\infty}^{\infty}dx\,x^n 
\exp[-\Lambda(\pm x^2-x^4)]\over\int_{-\infty}^{\infty} dx \,
\exp[-\Lambda(\pm x^2-x^4)]},
\label{eq3.8}
\end{equation}
where $\Lambda=m^4/g$ is a large positive parameter as $g\to0$ and we have
distinguished between the two cases $m^2>0$ and $m^2<0$.

\begin{figure}[p]
\centerline{\epsfig{file=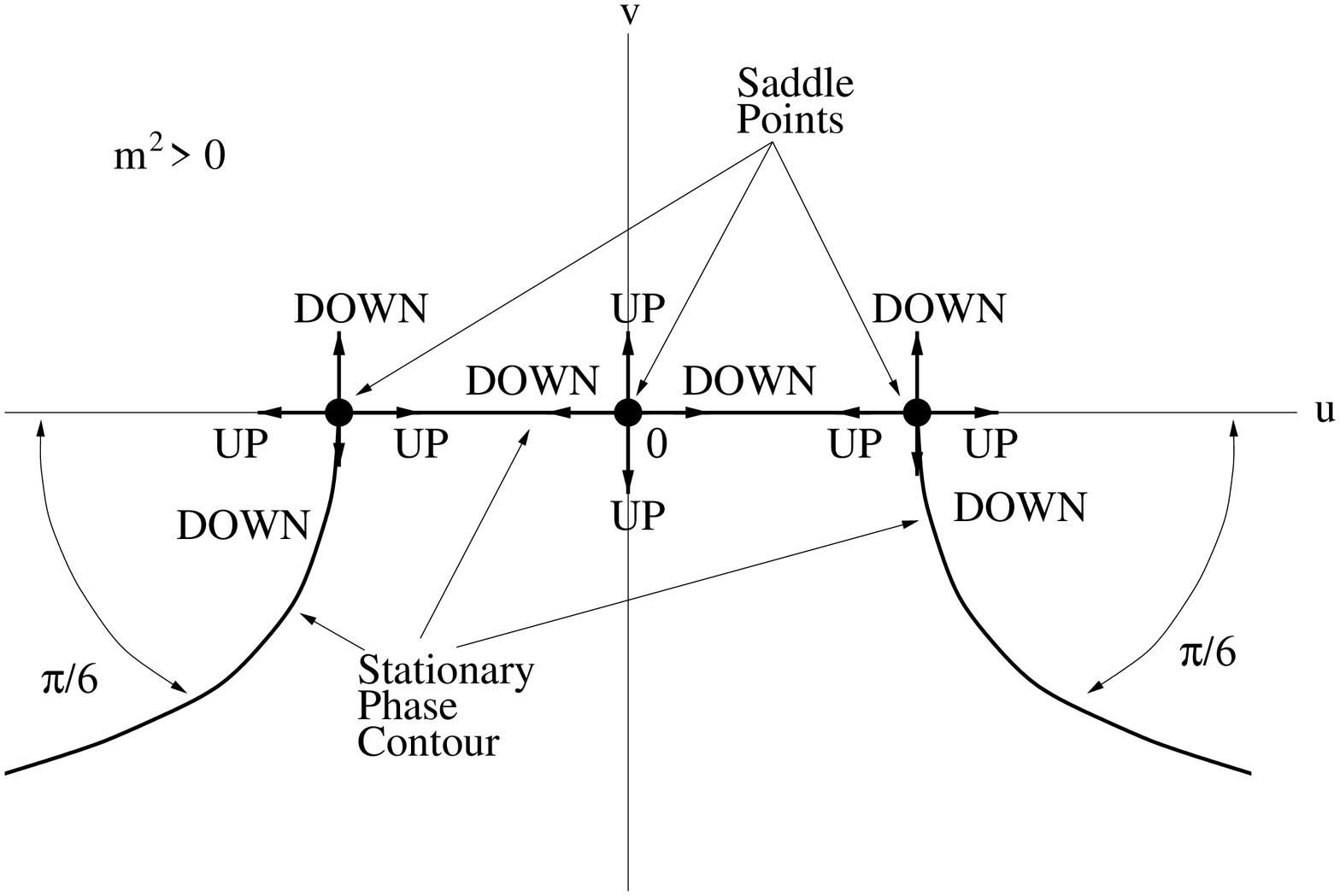,width=10cm}}
\caption{Saddle points and steepest-descent paths for the function $\rho(x)$
in Eq.~(\ref{eq3.9}) for the case $m^2>0$.}
\label{fig3}
\end{figure}

\begin{figure}[p]
\centerline{\epsfig{file=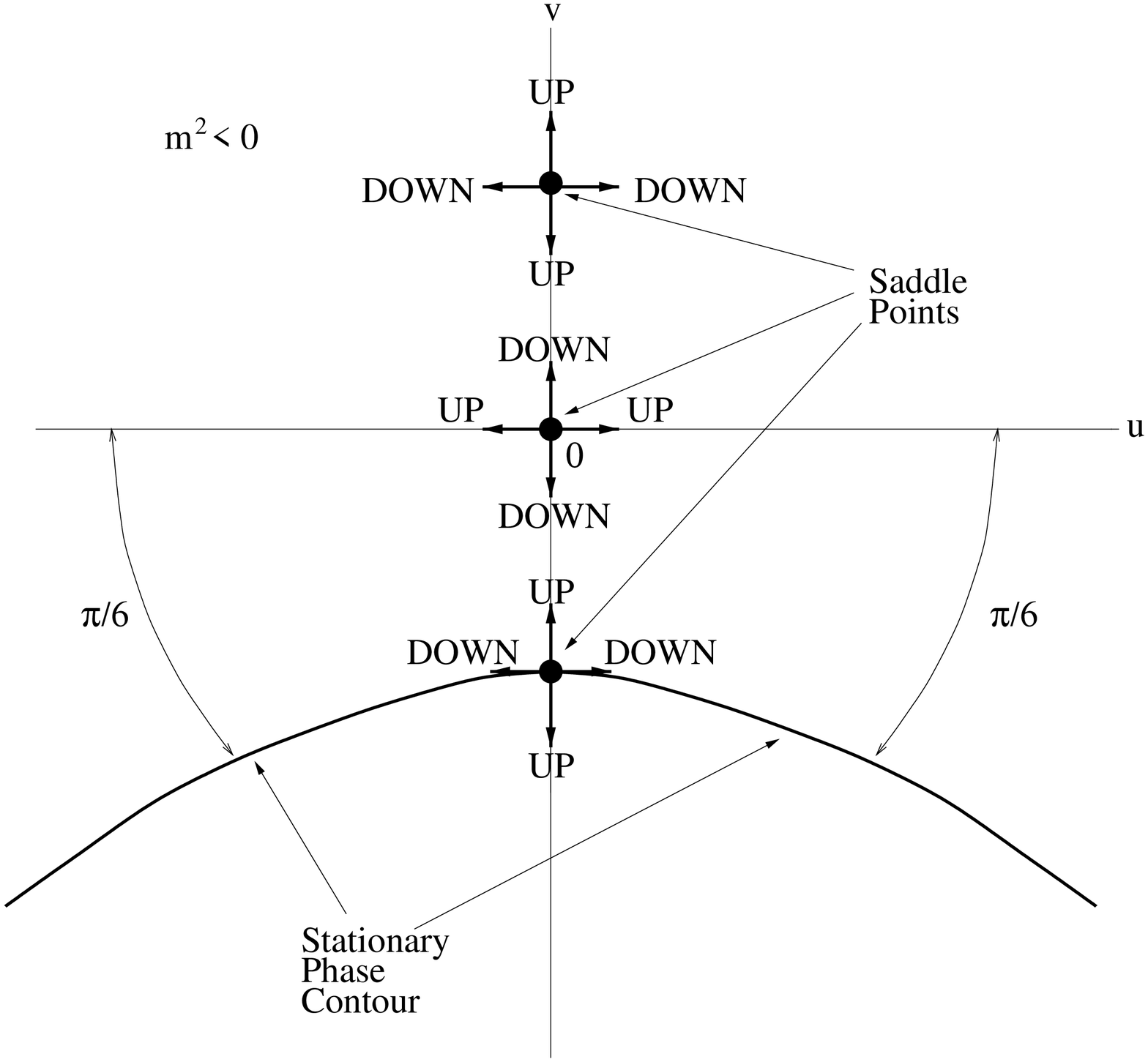,width=10cm}}
\caption{Saddle points and steepest-descent paths for the function $\rho(x)$
in Eq.~(\ref{eq3.9}) for the case $m^2<0$.}
\label{fig4}
\end{figure}

To calculate the integrals in Eq.~(\ref{eq3.7}) asymptotically, we find the
stationary points of the function $\rho(x)$ in the exponent, where
\begin{equation}
\rho(x)=\pm{x^2}-x^4.
\label{eq3.9}
\end{equation} 
Setting the derivative of this function equal to zero gives the solutions
$x=0,~\pm1/\sqrt{2}$ for $m^2>0$ and $x=0,~\pm i/\sqrt{2}$ for $m^2<0$. 
These stationary points are shown on Figs.~\ref{fig3} and \ref{fig4}.

Next, we find the paths of steepest ascent and descent that pass
through these stationary points. Substituting $x=u+iv$ into Eq.~(\ref{eq3.9})
and separating the real and imaginary parts yields
\begin{eqnarray}
\rho(u+iv)=\pm(u^2-v^2)-u^4-v^4+6u^2v^2
+ i[\pm 2uv-4uv(u^2-v^2)].
\label{eq3.10}
\end{eqnarray}  
Thus, the paths along which ${\rm Im}\,\rho=0$ are given by $u=0,~v\neq 0$;
$v=0,~u\neq 0$; and $u^2=v^2\pm1/2$, where the $\pm$ depends on the sign of
$m^2$. 

To determine whether each of these paths is a steepest ascent or descent path, 
we perform a local analysis of the paths at each of the saddle points. The
results of this analysis are displayed in Figs.~\ref{fig3} and \ref{fig4}. The
stationary phase contour is chosen such that the boundary conditions are
obeyed. Using Eq.~(\ref{eq1.3}), we choose the end points that lie below the
real axis at the angles $-\pi/6$ and $-5\pi/6$. For $m^2>0$ the contour follows
the path $u=-\sqrt{v^2+1/2}$ from the lower left quadrant of the complex-$x$
plane up to the stationary point at $-1/\sqrt{2}$, then goes along the real axis
from $-1/\sqrt{2}$ through the stationary points at the origin and at
$1/\sqrt{2}$, and finally leaves the point $1/\sqrt{2}$ and follows the path
$u=\sqrt{v^2+1/2}$ down to the lower right quadrant of the complex-$x$ plane.
For $m^2<0$ the contour follows the path $v=-\sqrt{u^2+1/2}$, which passes
through the stationary point at $-i/\sqrt{2}$.

Now, we determine the disconnected Green's functions $F_n$ for both the $m^2>0$
and $m^2<0$ cases by evaluating the integrals in Eq.~(\ref{eq3.8}) along the
appropriate stationary phase contours. For $m^2>0$ the Green's functions with
odd subscript vanish (by oddness) along the part of the contour that lies on the
real axis. Evaluating the integrals asymptotically along the remaining parts of
the contour gives an exponentially small result; apart from a multiplicative
constant
\begin{equation}
F_{2n+1}\sim g^{-n-1/2}e^{-\Lambda/4}\quad(g\to0^+).
\label{eq3.11}
\end{equation}    
The dominant contribution to Green's functions of even subscript comes from 
the portion of the contour along the real axis. Evaluating the integrals near
the saddle point at the origin, we find that the small-$g$ behavior of
these Green's functions is given by
\begin{equation}
F_{2n}\sim m^{-2n}\quad(g\to0^+),
\label{eq3.12}
\end{equation}    
apart from a multiplicative constant. Note that the Green's functions with even
subscript behave very differently than the Green's functions with odd subscript.
Further, notice that the behavior of the Green's functions with odd subscript is
inherently non-perturbative and, as a result, Feynman perturbative methods
cannot be used to calculate these Green's functions.

For $m^2<0$ the dominant contribution to {\it all} of the Green's functions
comes from the saddle point at $-i/\sqrt{2}$. As a result, {\it all} of the
Green's functions exhibit similar behavior. To be specific, we have
\begin{equation}
F_n\sim (-i)^n|m|^{n}g^{-n/2}\quad(g\to0^+).
\label{eq3.13}
\end{equation}     
Thus, once again, it is clear that these results cannot be obtained using
Feynman diagrams. As we will see in Sec.~\ref{s5} this theory is also the one
that is asymptotically free in four dimensions and the more interesting of
the two cases.

\section{Numerical Study of One-Dimensional Theories}
\label{s4}

Our success with zero-dimensional massless theories prompts us to study
one-dimensional (quantum mechanical) massless theories of the form in
Eq.~(\ref{eq1.4}). Numerical computations have been performed in one-dimension
which allow us to compare with exact numbers \cite{r1}. 

In analogy with the last section, we find the first approximation to $G_1$ and
$G_2$ by solving the one-dimensional massless versions of Eqs.~(\ref{eq2.28})
and (\ref{eq2.32}). Observe that with $m=0$ we obtain Eq.~(\ref{eq3.4}) once
again. In fact, Eq.~(\ref{eq3.4}) holds independent of the dimension. The second
Schwinger-Dyson equation is given by 
\begin{eqnarray}
-\partial^2G_2(x-y)+(N-1)g(iG_1)^{N-2}w_{N-2}(\gamma_0)G_2(x-y) 
 =\delta(x-y).
\label{eq4.1}
\end{eqnarray}  
This equation depends on the dimension $D$ through the partial derivative.
If we introduce the variable $M$ defined by $M^2=(N-1)g(iG_1)^{N-2}w_{N-2}(
\gamma_0)$, it is clear that Eq.~(\ref{eq4.1}) is just the equation for the
Feynman propagator, whose solution in one-dimension can be written
\begin{equation}
G_2(x-y)={1\over2M}e^{-|x-y|}.
\label{eq4.2}
\end{equation}
Consequently, $G_2(0)=1/(2M)$. Recalling the definition of $\gamma$ we use
$G_2(0) =-G_1^2/\gamma_0^2$ as we did in the previous section to obtain an
expression for $G_1$:
\begin{equation}
G_1=-i\left[{\gamma_0^4\over4(N-1)gw_{N-2}(\gamma_0)}\right]^{1/(N+2)}.
\label{eq4.3}
\end{equation}
Substituting $G_1$ into our expression for $M$ yields
\begin{equation}
M=\left[(N-1)gw_{N-2}(\gamma_0) \left({\gamma_0 \over \sqrt{2}}\right)^{N-2}
\right]^{2/(N+2)},
\label{eq4.4}
\end{equation}
which further allows us to write $G_2(0)$ as
\begin{equation}
G_2(0)={1\over2}\left[{1\over g(N-1)w_{N-2}(\gamma_0)}\left(\gamma_0\over
\sqrt{2}\right)^{2-N}\right]^{2/(N+2)}.
\label{eq4.5}
\end{equation}    
For field theories, $M$ represents the renormalized mass, which is nothing
more than the difference in energy between the first excited state and the
ground state $E_1-E_0$. Table \ref{t2} compares the exact values of $M$ and
$G_1$ with the corresponding first approximations for the cases $N=3$, $N=4$,
and $N=5$. (Here, we must set $g=N/2$ and multiply $M$ by $2$ to match the
Lagrangians studied in Ref.~\cite{r1}.)

In one dimension it is difficult to obtain the second approximation, even for a
specific $N$, because it requires solving coupled systems of nonlinear
differential equations. As in the previous section, the easiest case to study is
$N=3$. The first three Schwinger-Dyson equations are given in Sec.~\ref{s1} as
Eqs.~(\ref{eq2.10}), (\ref{eq2.12}), and (\ref{eq2.13}) with $m=0$. To close
this system of equations we set $G_4=0$ and $z=y$. Then, the third
Schwinger-Dyson equation becomes
\begin{eqnarray}
-\partial^2G_3(x-y)
+gi[2G_1G_3(x-y)+2G_2^2(x-y)]=0.
\label{eq4.6}
\end{eqnarray}

Next, we Fourier transform the second and third
Schwinger-Dyson equations to obtain
\begin{equation}
(p^2+M^2){\tilde G}_2(p)+gi{\tilde G}_3(p)=1
\label{eq4.7 }
\end{equation}
and
\begin{eqnarray}
(p^2+M^2){\tilde G}_3(p)
+2gi\int_{-\infty}^{\infty}{dq\over 2\pi} {\tilde G}_2(q)
{\tilde G}_2(p-q)=0,
\label{eq4.8}
\end{eqnarray}
where we have used the same definition of $M$ as above, the convolution property
of Fourier transforms, and the translation invariance of Green's functions
$G_3(0,x-y)=G_3(x-y,x-y)=G_3(x-y)$.

We now solve for $\tilde G_2(p)$ and obtain
\begin{eqnarray}
{\tilde G}_2(p)={1\over p^2+M^2}
-{2g^2\over(p^2+M^2)^2}\int_{-\infty}^{\infty} 
{dq\over2\pi}{\tilde G}_2(q){\tilde G}_2^{~}(p-q).
\label{eq4.9}
\end{eqnarray}

The simplest approach to solving this nonstandard integral equation is to
iterate it to high order in the small parameter $g$. This iterative
procedure can be represented in terms of diagrams. These diagrams all have a
similar structure: They begin with one line that branches into two lines.
This branching process continues until the maximum number of lines is
attained. Then the process is reversed, with lines combining in pairs until
only one line remains. We were able to perform the calculations symbolically on
a computer. We calculated the propagator to order ${\rm O}(g^{10})$; the
first three terms in the expansion are
\begin{eqnarray}
{\tilde G}_2(p)&=&{1\over p^2+M^2}-{2g^2\over M(p^2+M^2)^2(p^2+4M^2)^2}
\nonumber\\
&&\qquad+{(456M^4+70M^2p^2+4p^4)g^4\over9M^6(p^2+M^2)^2(p^2+4M^2)^2
(p^2+9M^2)}
+{\rm O}(g^6).
\label{eq4.10}
\end{eqnarray}

Now that we have ${\tilde G}_2(p)$ to high order in $g$, we make 
the {\it ansatz} that it can be expressed in the form
\begin{eqnarray}
{\tilde G}_2(p)
={Z_1\over p^2+M_1^2} +{Z_2\over p^2+M_2^2}+{Z_3\over p^2+M_3^2}+\ldots,
\label{eq4.11}
\end{eqnarray}  
where $M_n=nM+b_{1,n}g^2+b_{2,n}g^4+b_{3,n}g^6+\ldots$, 
$Z_1=1+a_{1,1}g^2+a_{2,1}g^4+a_{3,1}g^6+\ldots$,
$Z_2=a_{1,2}g^2+a_{2,2}g^4+a_{3,2}g^6+\ldots$,
$Z_3=a_{2,3}g^4+a_{3,3}g^6+\ldots$, and so on.
By matching this {\it ansatz} to our calculation, we determine the 
coefficients $a_{k,n}$ and $b_{k,n}$. [The expansion of the 
{\it ansatz} does not exactly match the expansion of our calculation.  
This is easily understood because our calculation for ${\tilde G}_2(p)$ only
involves special diagrams described above, while the {\it ansatz} involves all
types of diagrams. However, the system of equations is neither overdetermined
nor underdetermined, so all coefficients may still be calculated.]

The series for the $M_n$ are
\begin{eqnarray}
M_1&=&M+{g^2\over3M^4}-{31g^4\over72M^9}+{1279g^6\over1944M^{14}}
-{98287g^8\over93312M^{19}}+{9641179g^{10}\over5598720M^{24}},\nonumber\\
M_2&=&2M+{2g^2\over3M^4}-{11g^4\over108M^9}+{133g^6\over1944M^{14}}
+{33161g^8\over279936M^{19}},\nonumber\\
M_3&=&3M+{g^2\over M^4}+{29g^4\over216M^9}+{101g^6\over486M^{14}},\nonumber\\
M_4&=&4M+{4g^2\over3M^4}+{109g^4\over270M^9},\nonumber\\
M_5&=&5M+{5g^2\over3M^4}.  
\label{eq4.12}
\end{eqnarray}

Since we now have expressions for $M_n$ in terms of $M$, we need only
determine $M$ accurately to finish the calculation. This is done
using the first Schwinger-Dyson equation (\ref{eq2.10}), which implies that
\begin{equation}
M^4=4g^2G_2(0).
\label{eq4.13}
\end{equation}
Observe that
\begin{equation}
G_2(0)=\int_{-\infty}^{\infty}{dp\over 2\pi} {\tilde G}_2(p).
\label{eq4.14}
\end{equation}
So, based on our calculation of the first three terms above, the first few terms
are
\begin{equation}
G_2(0)={1\over2M} -{g^2\over9M^6}+{7g^4\over96M^{11}}.
\label{eq4.15}
\end{equation}
This allows us to express $M$ and $g$ as 
\begin{eqnarray}
M^2&=&\sqrt{4g^2G_2(0)}\nonumber\\
&=&\sqrt{{2g^2\over M}-{4g^4\over9M^6}+{7g^6\over24M^{11}}+{\rm O}(g^8)}.
\label{eq4.16}
\end{eqnarray}
Keeping terms to order ${\rm O}(g^{10})$ we obtain $M=1.126151g^{2/5}$.

With this value of $M$, the $M_n$ become numerical series multiplied by an
overall factor of $g^{2/5}$. Successive terms in this numerical series decrease
in magnitude and thus the series appears to be convergent. To compare with the
numerical results and match the Lagrangian in Ref.~\cite{r1}, we set $g=3/2$ and
multiply the $M_n$ by 2; results are given in Table \ref{t3}.

\section{Schwinger-Dyson Equations in $D$ Dimensions}
\label{s5}

In this section we show how to solve truncated systems of Schwinger-Dyson
equations in $D$ dimensions. First, we consider the case $D<2$, in which it
is not necessary to perform any renormalization. Then we consider the case
of arbitrary $D$, in which it is necessary to discuss renormalization.

\subsection{Schwinger-Dyson Equations for $D<2$}
\label{ss5a}

In this subsection we solve Eqs.~(\ref{eq2.28}) and (\ref{eq2.32}) in arbitrary
dimension $D$ with $m=0$. This calculation is a straightforward generalization
of the one for $D=1$.

As previously stated, when $m=0$, Eq.~(\ref{eq2.28}) implies Eq.~(\ref{eq3.4}).
In addition Eq.~(\ref{eq2.32}) continues to imply Eq.~(\ref{eq4.1}),
\begin{equation}
-\partial^2 G_2(x-y)+M^2 G_2(x-y)=\delta(x-y),
\label{eq5.02}
\end{equation}
in which we have defined the renormalized mass $M^2$ by
\begin{equation}
M^2=-gi^N (N-1) G_1^{N-2} w_{N-2}(\gamma_0).
\label{eq5.03}
\end{equation}
Equation (\ref{eq5.02}) is just the differential equation satisfied by the
Feynman propagator.  

We solve these two equations in arbitrary dimension $D$ by taking the
Fourier transform to obtain the propagator:
\begin{equation}
{\tilde G}_2(p)=1/(p^2+M^2).
\label{eq5.04}
\end{equation}
Fourier transforming this propagator back to position space and then setting
$x=y$ gives 
\begin{eqnarray}
G_2(0)=M^{D-2}\Gamma(1-D/2)\pi^{-D/2}2^{-D}.
\label{eq5.05}
\end{eqnarray}
Using Eqs.~(\ref{eq2.25}) and (\ref{eq5.03}) we solve for $G_1$:
\begin{eqnarray}
G_1=-i\left\{[(N-1)gw_{N-2}(\gamma_0)]^{(D-2)/2}\Gamma(1-D/2)
(4\pi)^{-D/2}\gamma_0^2\right\}^{2/(-ND+2N+2D)}.
\label{eq5.06}
\end{eqnarray}
Substituting this result into Eq.~(\ref{eq5.03}) yields
\begin{eqnarray}
M=\Bigm\{[(N-1)gw_{N-2}(\gamma_0)]^2\Bigm[\Gamma(1-D/2)
(4\pi)^{-D/2} \gamma_0^2\Bigm]^{N-2}\Bigm\}^{1/(-ND+2N+2D)},
\label{eq5.07}
\end{eqnarray}
and substituting the last two results into Eq.~(\ref{eq5.05}) gives
an expression for $G_2$. These expressions reduce to the zero-dimensional
and one-dimensional solutions given in the previous two sections.
Notice that each of these expressions becomes singular at $D=2$.

Using these approximate solutions we can determine
the large-$N$ behavior of the Greens functions as $N\to\infty$. To do so,
we need to determine the asymptotic behavior of the zeros of the parabolic
cylinder function. According to Ref.~\cite{Szego}, the largest zero of
$w_{n-1}(\gamma)$ for large $n$ is given by $\gamma_0\sim2\sqrt{n}$.
Substituting this into the integral representation for $w_{n-2}(\gamma)$ and
using Eq.~(\ref{eq2.31}) we perform a steepest-descent calculation to obtain 
$w_{n-2}\sim\sqrt{2/\pi}3^{-1/6}\Gamma(1/3)2^{-n}n^{1/6}e^{n/2}$. We have
verified these results numerically to high accuracy. Taking $N\to\infty$ in the
expressions for $G_1$, $G_2$, and $M$ above and using the asymptotic
results for the parabolic cylinder function, we obtain
\begin{eqnarray}
G_1&\sim&-2i/\sqrt{e}=-1.21306i,\nonumber\\
G_2&\sim&1/(Ne),\nonumber\\
M^2&\sim&\left[e\Gamma(1-D/2)(4\pi)^{-D/2}N\right]^{2/(2-D)}.
\label{eq5.08}
\end{eqnarray}
Observe that $G_1$ is independent of both $N$ and $D$ to this order, $G_2$
depends on $N$ but not on $D$, and $M^2$ depends on both $N$ and $D$. These
results are valid for large $N$ for all $D<2$. These properties are evident
in zero dimensions from Eqs.~(\ref{eq3.2}) and (\ref{eq3.3}). While the behavior
of $G_1$ and $\langle0|\phi^2(0)|0\rangle$ is correctly predicted, the 
behavior of $G_2$ is not because the first-order behaviors of $G_1$ and 
$\langle0|\phi^2(0)|0\rangle$ cancel, leaving a second-order term to describe
$G_2$. Also, this calculation predicts that $M^2$ increases like $N^{2/(2-D)}$.
That is, for large $N$ the separation between the energy levels diverges and
hence, the energy levels must diverge. In $D=1$ our approximation suggests that
$M$ grows like $N$. In fact, $M$ grows like $N^2$ in $D=1$ as discussed in
Ref.~\cite{r9}. This is the simplest possible truncation, but it suggests the
correct behavior.   

\subsection{Perturbative Renormalization of the Schwinger-Dyson Equations in
$D$-Dimensions and Leading-Order Calculation of the Beta Function}
\label{ss5b}

Let us first consider the ${\cal PT}$-symmetric $ig\phi^3$ theory in six
space-time dimensions. The Green's functions for that theory are governed by the
system of equations beginning with Eq.~(\ref{eq2.10}), (\ref{eq2.12}), and
(\ref{eq2.13}). Let us seek a perturbative solution to this system of equations
in which $G_n\sim g^{n-2}$. In leading order we have first, from
Eq.~(\ref{eq2.10}), 
\begin{eqnarray}
G_1=i{m^2\over g}.
\label{e5.1}
\end{eqnarray}
Then, from Eq.~(\ref{eq2.12}), the two-point function in momentum-space is
\begin{eqnarray}
\tilde G_2(p)=\int d^6x \,e^{ip(x-y)}G_2(x-y)={1\over p^2+M^2},
\label{e5.2}
\end{eqnarray}
where
\begin{eqnarray}
M^2=m^2+2igG_1=-m^2.
\label{e5.3}
\end{eqnarray}
Thus, in order to avoid unphysical singularities, we must have for this type of
solution $m^2<0$. The leading solution to Eq.~(\ref{eq2.13}) is
\begin{eqnarray}
\tilde G_3(p,q)={2ig\over(p^2+M^2)(q^2+M^2)[(p+q)^2+M^2]},
\label{e5.4}
\end{eqnarray}
which has an obvious interpretation as a vertex with three external lines.

More generally, the solution to Eq.~(\ref{eq2.10}) is
\begin{eqnarray}
G_1={1\over2}\left[im^2/g\pm i\sqrt{\left(m^2/g\right)^2+4G_2(0)}\right],
\label{e5.5}
\end{eqnarray}
which corresponds in the perturbative case, where $|G_2(0)|\ll(m^2/g)^2$, to
\begin{eqnarray}
G_1=\left\{
\begin{array}{c}
i{m^2\over g},\\ -i{g\over m^2}G_2(0).
\end{array}\right.
\label{e5.6}
\end{eqnarray}
The second solution given in Eq.~(\ref{e5.6}) corresponds to the usual
perturbative tadpole contribution to the vacuum expectation value of the field,
while the first is the new, nontrivial solution given in Eq.~(\ref{e5.1}).

Perturbatively solving the next in the sequence of Schwinger-Dyson equations, we
obtain for the four-point function in leading order
\begin{eqnarray}
\tilde G_4(p,q,r)
&=&{-2ig\over((p+q+r)^2+M^2)(p^2+M^2)(q^2+M^2)(r^2+M^2)}\nonumber\\
&&\times\left[{1\over(p+q)^2+M^2}+{1\over(p+r)^2+M^2}+{1\over(q+r)^2+M^2}
\right].\nonumber\\
\label{e5.7}
\end{eqnarray}
Inserting this back into Eq.~(\ref{eq2.13}), we obtain the one-loop correction
to the three-point function. Apart from a tadpole term, this is just the same as
that found in the conventional $\phi^3$ theory; correspondingly, the beta
function is obtained from that in the conventional theory by the replacement
$g\to ig$ (see, for example, Ref.~\cite{muta})
\begin{eqnarray}
\beta(g)=\mu {\partial g\over\partial\mu}={3\over2}\left({g\over4\pi}\right)^3,
\label{e5.8}
\end{eqnarray}
where $g(\mu)$ is the running coupling at scale $\mu$. Unlike the usual
$\phi^3_6$ theory, the ${\cal PT}$ symmetric theory given here is not
asymptotically free.

Of course, the $\phi_4^4$ theory is of far greater interest. In particular, it
plays a crucial role in the standard model as the origin of particle masses
through the Higgs mechanism.  Yet the triviality of that theory is a source of
difficulty. What happens here, when we set $N=4$ in Eq.~(\ref{eq1.1})?

The first few Schwinger-Dyson equations are given in  
Eqs.~(\ref{eq2.16}), (\ref{eq2.18}), (\ref{eq2.19}), and (\ref{eq2.20}). Note
that the last three equations can be simplified through the introduction of the
renormalized mass
\begin{eqnarray}
M^2=m^2-3gG_2(0)-3gG_1^2.  
\label{e5.11}
\end{eqnarray}
Thus we obtain the following equations for the two-point function,
\begin{eqnarray}
(-\partial^2+M^2)G_2(x-y)-gG_1G_3(0,x-y)
-gG_4(0,0,x-y)=\delta(x-y),
\label{e5.10}
\end{eqnarray}
the three-point function
\begin{eqnarray}
&&(-\partial^2+M^2)G_3(x-y,x-z)
-6gG_1G_2(x-y)G_2(x-z)-3g[G_2(x-y)G_3(0,x-z)\nonumber\\
&&+G_2(x-z)G_3(0,x-y)]
-3gG_1G_4(0,x-y,x-z)-gG_5(0,0,x-y,x-z)=0,
\label{e5.12}
\end{eqnarray}
and the four-point function
\begin{eqnarray}
&&(-\partial^2+M^2)G_4(x-y,x-z,x-w)
-6gG_2(x-y)G_2(x-z)G_2(x-w)\nonumber\\
&&\mbox{}-3g[G_2(x-y)G_4(0,x-z,x-w)
+G_2(x-z)G_4(0,x-y,x-w)\nonumber\\
&&\mbox{}+G_2(x-w)G_4(0,x-y,x-z)]
-6gG_1[G_2(x-y)G_3(x-z,x-w)\nonumber\\
&&\mbox{}+G_2(x-z)G_3(x-y,x-w)
+G_2(x-w)G_3(x-y,x-z)]\nonumber\\
&&\mbox{}-3g[G_3(0,x-y)G_3(x-z,x-w)
+G_3(0,x-z)G_3(x-y,x-w)\nonumber\\
&&\mbox{}+G_3(0,x-w)G_3(x-y,x-z)]
-3gG_1G_5(0,x-y,x-z,x-w)\nonumber\\
&&\mbox{}-3gG_6(0,0,x-y,x-z,x-w)=0.
\label{e5.13}
\end{eqnarray}

Now there are two regimes. If $m^2>0$ the only consistent perturbative solution
to the above system of equations is one in which the odd Green's functions are
exponentially small, $G_{2n+1}\sim e^{-1/g}$, and the even and odd Green's
functions decouple. [This result is analogous to that in Eq.~(\ref{eq3.11}).]
The even Green's functions possess the same perturbative
expansion as in the usual $\phi^4$ theory except for a change of sign of the
coupling constant, so again the sign of the beta function reverses:
\begin{eqnarray}
\beta(g)=-27\left(g\over2\pi\right)^2.
\label{e5.14}
\end{eqnarray}
This theory is asymptotically free. Furthermore, in the nonperturbative regime
it exhibits parity symmetry breaking, but possesses ${\cal PT}$ symmetry because
$G_1$ is imaginary.

The other regime is even more interesting. If $m^2<0$ it is consistent to
proceed in analogy with our treatment of the $\phi^3$ theory above. We may
assume a perturbative solution of the form $G_n\sim g^{n/2-1}$,
as we have already seen in Eq.~(\ref{eq3.13}). Then we have a
purely imaginary vacuum expectation value of the field, from Eq.~(\ref{eq2.16}):
\begin{eqnarray}
G_1=\sqrt{m^2\over g}, 
\label{e5.15}
\end{eqnarray}
while the leading two-point function has the usual form of a propagator:
\begin{eqnarray}
\tilde G_2(p)={1\over p^2+M^2}.
\label{e5.16}
\end{eqnarray}
Here the renormalized mass in leading order is positive:
\begin{eqnarray}
M^2=m^2-3gG_1^2=-2m^2.
\label{e5.17}
\end{eqnarray}
The leading three-point function has an evident diagrammatic interpretation:
\begin{eqnarray}
\tilde G_3(p,q)=6gG_1{1\over p^2+M^2}{1\over q^2+M^2}{1\over(p+q)^2+M^2}.
\label{e5.18}
\end{eqnarray}
The tree-level four-point function is easily extracted from Eq.~(\ref{e5.13}):
\begin{eqnarray}
\tilde G_4(p,q,r)
&=&{6g\over(p^2+M^2)(q^2+M^2)(r^2+M^2)[(p+q+r)^2+M^2]}\nonumber\\
&&\times\bigg[1+{3m^2\over(p+q)^2+M^2}
+{3m^2\over(p+r)^2+M^2}+{3m^2\over(q+r)^2+M^2}\bigg],
\label{e5.19}
\end{eqnarray}
being composed of contibutions from primitive four-point and three-point
vertices.

\begin{figure}[p]
\centerline{\epsfig{file=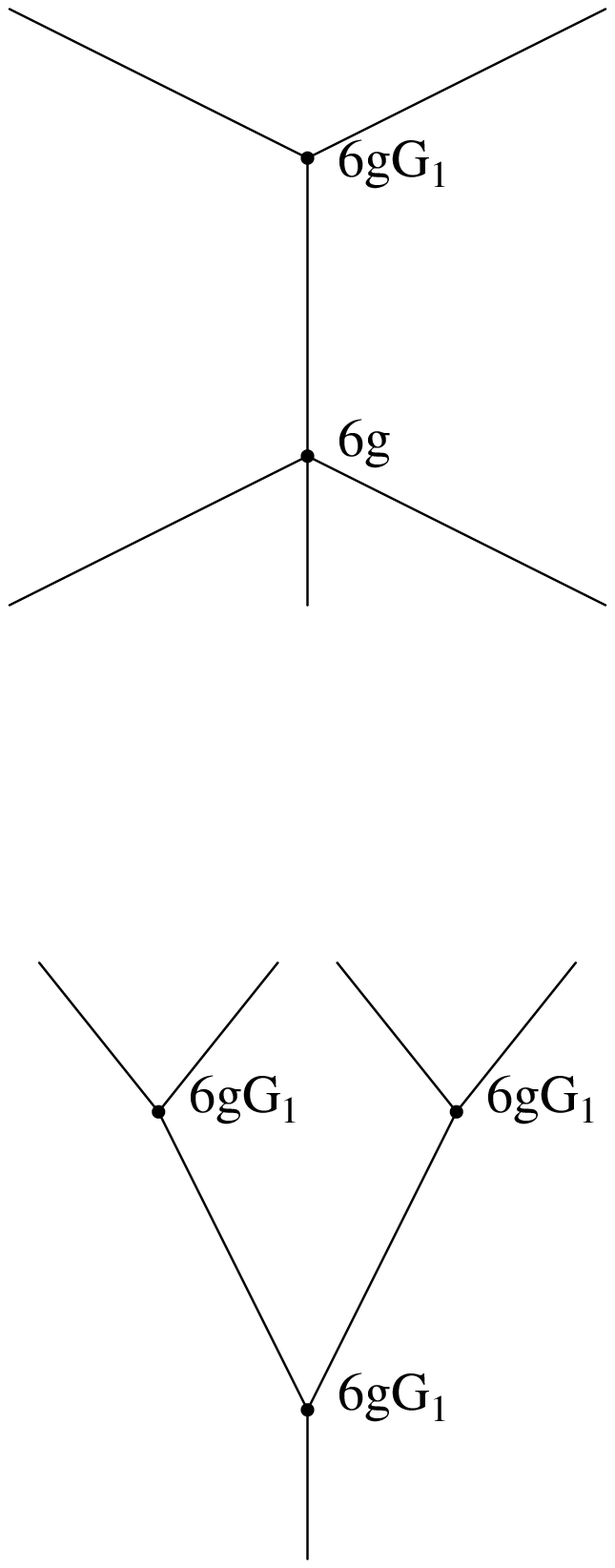,width=5cm}}
\caption{Lowest order graphs contributing to $G_5$.}
\label{fig5}
\end{figure}

Now we have perturbative parity symmetry breaking: the scalar field acquires a
vacuum expectation value comparable to that of the gauge bosons in the standard
model, $\sqrt{g}G_1$. Further, it appears likely that the theory is
asymptotically free, because the sign of the four-point vertex is reversed.
Indeed, apart from one-particle-reducible graphs, Eq.~(\ref{e5.19}) gives just
the usual primitive vertex in the high momentum limit, except for a change
in sign. The
theory is renormalizable because, apart from divergences associated with the 2-,
3-, and 4-point functions, no additional divergences occur. This is due to the
fact that, for example, the 5-point function has no primitive vertices, as can
be easily seen from the next in the sequence of Schwinger-Dyson equations after 
Eq.~(\ref{e5.12}). The lowest order diagrams contributing to $G_5$ are as
sketched in Fig.~\ref{fig5}.

We have shown that the signs of the beta functions for a conventional $g\phi^3$
theory and for a ${\cal PT}$-symmetric $ig\phi^3$ theory in six space-time
dimensions are reversed. Thus, while the former theory is asymptotically
free, the latter is not. Similarly, the beta functions for a conventional
$g\phi^4$ theory and for a ${\cal PT}$-symmetric $-g\phi^4$ theory in four
space-time dimensions are reversed. Thus, while the former theory is not
asymptotically free, the latter is. Similarly, as we have already argued
in Ref.~\cite{r10}, we believe that while conventional quantum
electrodynamics is not asymptotically free, ${\cal PT}$-symmetric
electrodynamics is asymptotically free and possesses a nontrivial fixed point.

One of us, CMB, thanks Arthur Lue for many useful discussions at the Aspen
Center for Physics.
This work was supported in part by the U.S. Department of Energy.

\begin{table}
\caption[t1]{Exact values $iG_1^{\rm exact}g^{1/N}$ and $G_2^{\rm exact}
g^{2/N}$ [see Eqs.~(\ref{eq3.2}) and (\ref{eq3.3})] compared with the first
approximations $iG_1^{\rm SD}g^{1/N}$ and $G_2^{\rm SD}g^{2/N}$, which are
obtained from the first two Schwinger-Dyson equations and given in
Eqs.~(\ref{eq3.5}) and (\ref{eq3.6}). This is done for the three cases $N=3$,
$N=4$, and $N=5$. Observe that the percent error increases with $N$ for
$G_1$ and decreases with $N$ for $G_2$. These trends continue until
$G_1$ has a maximum error of $23.2\%$ at $N=53$, and until $G_2$ has a minimum
error of $0.34\%$ at $N=8$. For $N>53$ the error for $G_1$ decreases until it
levels off at $21.3\%$. For $N>8$ the error for $G_2$ increases like $N$. The
large-$N$ behavior of our approximations is more fully discussed later.}
\begin{tabular}{ldddd}
$N$ & $iG_1^{\rm exact}g^{1/N}$ & $iG_1^{\rm SD}g^{1/N}$ & $G_2^{\rm exact}
g^{2/N}$ & $G_2^{\rm SD}g^{2/N}$ \\ \tableline
3 & 0.72901 & 0.79370 & 0.53146 & 0.27516\\
4 & 0.97774 & 1.10668 & 0.28000 & 0.14907\\
5 & 1.07865 & 1.24829 & 0.16433 & 0.10158\\
\end{tabular}
\label{t1}
\end{table}

\begin{table}
\caption[t2]{Exact values of $iG_1^{\rm exact}$ and $M^{\rm exact}=E_1-E_0$ (see
Ref.~\cite{r1}) compared with the first approximations $iG_1^{\rm SD}$ and
$M^{\rm SD}$, which are obtained from the first two Schwinger-Dyson equations
(\ref{eq4.3}) and (\ref{eq4.4}). This is done for the three cases $N=3$,
$N=4$, and $N=5$. Note that the error is worse than for the zero-dimensional
theories, and that the error for $M$ increases with $N$. However, the error for
$iG_1$ is smallest for $N=4$.}
\begin{tabular}{ldddd}
$N$ & $iG_1^{\rm exact}$ & $iG_1^{\rm SD}$ & $M^{\rm exact}$ & $M^{\rm SD}$
\\ \tableline
3 & 0.59007 & 0.37011 & 2.95293 & 2.70192\\
4 & 0.86686 & 0.82548 & 4.52620 & 3.63424\\
5 & 1.01310 & 1.15416 & 6.70000 & 4.72160\\
\end{tabular}
\label{t2}
\end{table}

\begin{table}
\caption[t3]{}
Schwinger-Dyson approximations for $M_n$ compared with exact values of the 
energy difference, $E_n-E_0$, calculated in Ref.~\cite{r1}. These approximations
are based on the truncation of the first three Schwinger-Dyson equations for an
$ig\phi^3$ field theory of the type in Eq.~(\ref{eq1.4}). Notice that $M_1$ is
greater than the numerical result in this case, while in Table \ref{t2}, $M_1$
was less than the numerical answer. This suggests that in one dimension the
oscillatory nature of successive approximations is present once again. Moreover,
the percent error has decreased significantly; the error is $8.50\%$ for $M_1$
calculated using the first two Schwinger-Dyson equations compared with $3.52\%$
calculated using the first three Schwinger-Dyson equations.    
\begin{tabular}{lddd}
$n$ & $E_n-E_0$ & $M_n^{SD}$ & \% error
\\ \tableline
1  & 2.952962   & 3.056763  & 3.52 \\
2  & 6.406007   & 6.191828  & 3.34\\
3  & 10.158155  & 9.610135  & 5.39\\
4  & 14.135286  & 12.871434 & 8.94\\
5  & 18.295263  & 15.681836 & 14.28\\
\end{tabular}
\label{t3}
\end{table}
\end{document}